\documentclass[11pt,a4paper]{article}
\pdfoutput=1
\usepackage[utf8]{inputenc}
\usepackage[english]{babel}
\usepackage{extarrows}
\usepackage{amsmath}
\usepackage{amsfonts}
\usepackage{amssymb}
\usepackage{graphicx}
\usepackage{fourier}

\usepackage{caption}
\usepackage{subcaption}
\usepackage{float}
\usepackage{color}
\usepackage{abstract}
\usepackage{appendix}
\usepackage{authblk}

\numberwithin{equation}{section}

\usepackage[hidelinks]{hyperref}

\newcommand\be{\begin{equation}}
\newcommand\ee{\end{equation}}
\newcommand\ba{\begin{eqnarray}}    
\newcommand\ea{\end{eqnarray}}      

\title{R\'{e}nyi entropies and area operator from gravity\\ with Hayward term}

\author[a]{Marcelo Botta-Cantcheff}
\author[a]{Pedro J. Martinez \footnote{martinezp@fisica.unlp.edu.ar} }
\author[a]{Juan F. Zarate}

\affil[a]{Instituto de F\'isica La Plata - CONICET and 

Departamento de F\'isica, Universidad Nacional de La Plata 

C.C. 67, 1900, La Plata, Argentina}

\date{}
\usepackage[left=2cm,right=2cm,top=2cm,bottom=2cm]{geometry}

\begin{document}
  
\maketitle
\thispagestyle{empty}

\begin{abstract}

In the context of the holographic duality, the entanglement entropy of ordinary QFT in a subregion in the boundary is given by a quarter of the area of an minimal surface embedded in the bulk spacetime. This rule has been also extended to a suitable one-parameter generalization of the von-Neuman entropy $\hat{S}_n$ that is related to the Rényi entropies $S_n$, as given by the area of a \emph{cosmic brane} minimally coupled with gravity, with a tension related to $n$ that vanishes as $n\to1$, and moreover, this parameter can be analytically extended to arbitrary real values. However, the brane action plays no role in the duality and cannot be considered a part of the theory of gravity, thus it is used as an auxiliary tool to find the correct background geometry.

In this work we study the construction of the gravitational (reduced) density matrix from holographic states, whose wave-functionals are described as euclidean path integrals with arbitrary conditions on the asymptotic boundaries, and argue that in general, a non-trivial Hayward term must be haven into account.
So we propose that the gravity model with a coupled Nambu-Goto action is not an artificial tool to account for the Rényi entropies, but it is present in the own gravity action through a Hayward term.
As a result we show that the computations using replicas simplify considerably and we recover the holographic prescriptions for the measures of entanglement entropy; in particular, derive an area law for the original R\'enyi entropies ($S_n$) related to a minimal surface in the $n$ replicated spacetime.
Moreover, we show that the gravitational modular flow contains the area operator and can explain the Jafferis-Lewkowycz-Maldacena-Suh proposal. 

\end{abstract}

\newpage

\section{Introduction}

The von Neumann entropy measures the entanglement of a physical system in a given state and for a specific subset of degrees of freedom, and 
the celebrated Ryu-Takayanagi (RT) \cite{RT} formula is a powerful tool to compute it in quantum field theory in the mindset of the gauge/gravity correspondence. This generalizes the Bekenstein-Hawking law for the thermodynamic entropy of Black holes \cite{BH}, and the entropy is given by a quarter of the area of the minimal surface embedded in the dual higher dimensional spacetime with gravity.
Since its discovery, a lot evidence of its validity had been collected, and it was finally been 
derived by computing the gravitational entropy with different replica methods \cite{Fursaev,Headrick,MaldaAitor}.

The Rényi entropies are a generalization of the von Neumann entropy labeled by an integer parameter $n$ \cite{Renyi},
 \be\label{defirenyi} S_n \equiv \frac{1}{1-n} \log{ \text{Tr} \rho^n }\ee
such that the standard von Neumann entropy $S\equiv-\text{Tr}\rho\log \rho$ is recovered in the limit $n\to 1$. There is an alternative family of measures of entanglement entropy related to the Rényi entropies, given by 
\be\label{defirenyi-hat} \hat{S}_n \equiv -n^2 \partial_n \left(\frac{1}{n} \log{ \text{Tr} \rho^n }\right)\ee
that also coincides with the von Neumann entropy as $n\to 1$, and has a very similar thermodynamic interpretation \cite{RenyiMod}. A similar area-law prescription for these entropies has been provided \cite{Dong16}, but in this case the extremal surface interacts with the background spacetime through a tension that depends on the parameter in the specific way
\be\label{tension-dongC}
T_n = \frac{n-1}{4n\,G}
\ee 
where $G$ is the Newton's constant.
In a computational sense, Rényi entropies are generally easier to handle. However, they are objects of interest in their own as they should provide a full understanding of the entanglement structure of the quantum state \cite{RenyiTeor,Headrick} and are known sometimes to be directly measured \cite{RenyiMedida}.
Rényi entropies have been previously studied in the holographic context \cite{Fursaev,Headrick,MaldaAitor,Dong16}. 

The proposal of \cite{Dong16} consists of an elegant Nambu-Goto action describing a cosmic brane coupled to gravity, with a tension that depends on $n$ and vanishes for $n=1$, such that the RT law is recovered. By virtue of \eqref{tension-dongC}, the parameter $n$ can be analytically extended to any real value. However, the origin of such brane is hard to be justified from standard holographic recipes. It cannot be argued in the own gravity theory, and need to be put by hand as an
tool to obtain the conical dominant solutions and explain the entropies \eqref{tension-dongC}. In other words, the problem is that the reduced density matrix should be calculated from a pure global state in gravity (in the Hartle-Hawking formalism), by tracing out the complementary dof's in the bulk theory but it does not explain the cosmic brane term or its effect in the solutions. 
In an alternative approach, specific states were considered with definite (extremal) area \cite{Dong18}, which provides a $S_n$ proportional to this area and independent on $n$. 

In this work we start from a different point of view that also captures the results for the entropy, and moreover, explains the cosmic brane contribution with the appropriate tension from the own theory of gravity, through a very plausible assumption on the correspondence between subsystems in both sides of the gravity/gauge duality.

\begin{figure}
  \centering
   \includegraphics[width=0.35\textwidth]{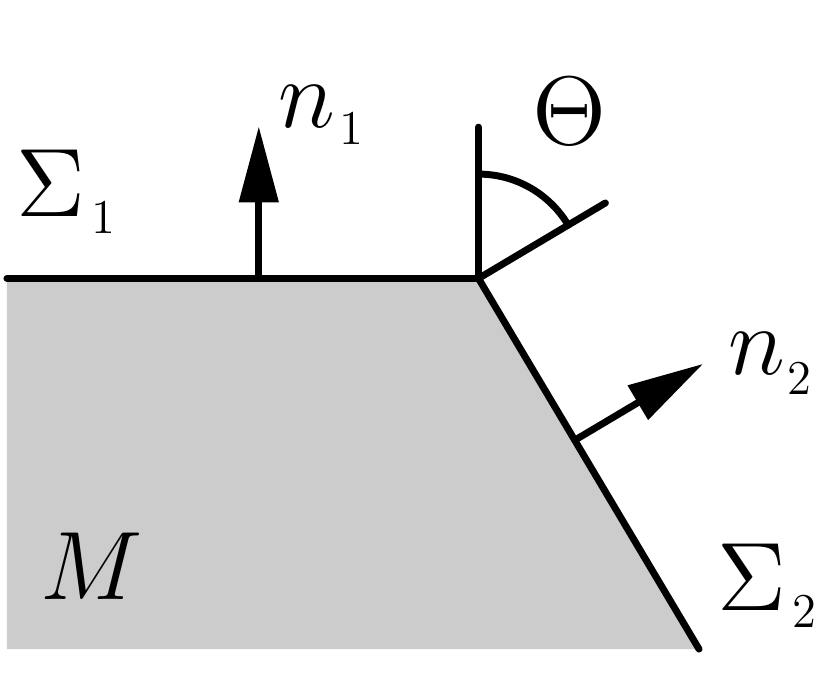}
  \caption{The figure shows the standard situation where a Hayward boundary term must be considered. The spacetime $M$ has a boundary with a non-smooth corner at $\Gamma=\Sigma_1\cap\Sigma_2$, which is described in terms of the angle $\Theta=\cos^{-1}\, (n_1 \cdot  n_2)$.} 
  \label{fig:H2}
\end{figure}

 The idea is to consider a generalization of the Gibbons Hawking boundary term to cases where the spacetime has a non smooth boundary proposed by Hayward in the 90's \cite{Hayward93}. For instance, if there is a co-dimension 2 corner $\Gamma$ (see fig. \ref{fig:H2}) that splits the spacetime boundary in two smooth components $\Sigma_{1,2}$ with respective normal vectors $n_{1,2}$, thus the standard gravitational action has an extra term given by
\be
\frac{1}{8\pi G}\int_\Gamma \; \cos^{-1}\, (n_1 \cdot  n_2) \;\sqrt{\gamma}  
\ee
where $\gamma$ is the induced metric on $\Gamma$. Since the boundary is fixed previously, the corner angle is arbitrarily fixed and the Hayward term is required to get a well posed variational problem.

In a very recent article \cite{Takayanagi19} Takayanagi and Tamaoka drew attention to a possible application of this term to holographic context and to the study of the entanglement entropy; in particular, using a replica trick calculation close to the Fursaev's approach \cite{Fursaev}, they showed that the von Neumann entanglement entropy can be explained by considering the Hayward term in the gravitational action.
The aim of the present work is to generalize this result to the Rényi entropies, precisely by showing that the cosmic brane term is explained in the own gravitational theory through a Hayward term. The presence of the Newton's constant $G$ in \eqref{tension-dongC} enforces such point of view.

Another important aspect captured by the present study is the modular Hamiltonian associated to the modular flow in gravity \cite{ModFlow,Haag}, which can be obtained from the gravitational density matrix that will include the area operator.
The area operator in a holographic context had been essayed in Ref. \cite{ensayoarea} as the gravity dual of the modular Hamiltonian in the gauge theory, and it could lead to the quantization of areas, at least in certain specific contexts such as black holes. 
Then a more detailed analysis of the area operator in a suitable (holographic) quantum gravity was provided in \cite{Jafferis}, and 
finally based on it, the presence of this area operator as part of the modular Hamiltonian of gravity was formulated  in Ref. \cite{JLMS}, in what is known as the JLMS conjecture.  In the present work, a precise (path integral) definition of area operator and how its matrix elements in a basis of bulk fields (i.e, boundary data on two copies of the entanglement wedge) can be computed in the semi-classical (large $N$) regime, will be obtained as a result. Moreover it will be shown that it is present in the gravitational modular Hamiltonian in agreement with the proposal JLMS proposal. 

This work is organized as follows. In Sec 2 we discuss how bipartite systems in the boundary QFT should be related to the possible partitions of the gravitational d.o.f. and give a plausible holographic prescription. In Sec 3 we describe the states and wave functional in gravity, and show how the Hayward term appear as more general (non smooth) initial surfaces are considered. In Sec 4 we describe the density matrix in gravity and show that the area operator appears. Sec. 5 is devoted to derive the area law for the von Neumann entropies with the Hayward term, and in Sec. 6 we generalize it using replicas and obtain the prescriptions for the Rényi and modified Rényi entropies. Finally, in Sec. 7 we study the modular flow in gravity (with Hayward term) and obtain the JLMS formula, properly involving the area operator. Concluding remarks are collected in Sec. 8. 

\begin{figure*}
\begin{subfigure}{0.49\textwidth}\centering
\includegraphics[width=.9\linewidth] {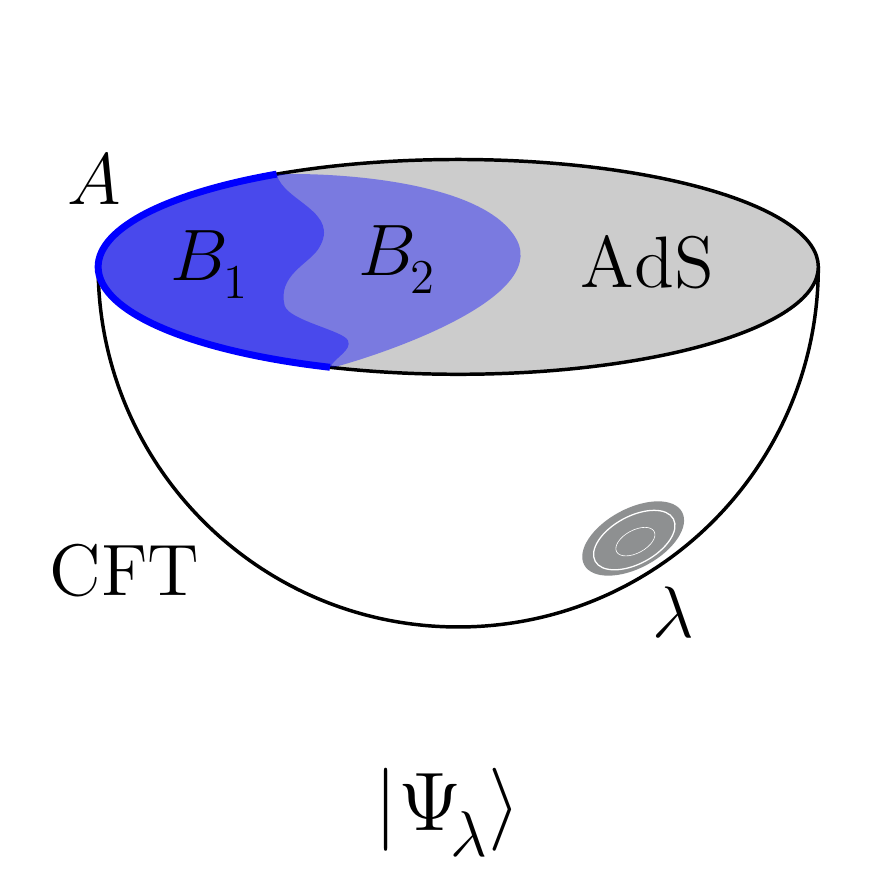}
\caption{}
\end{subfigure}
\begin{subfigure}{0.49\textwidth}\centering
\includegraphics[width=.9\linewidth] {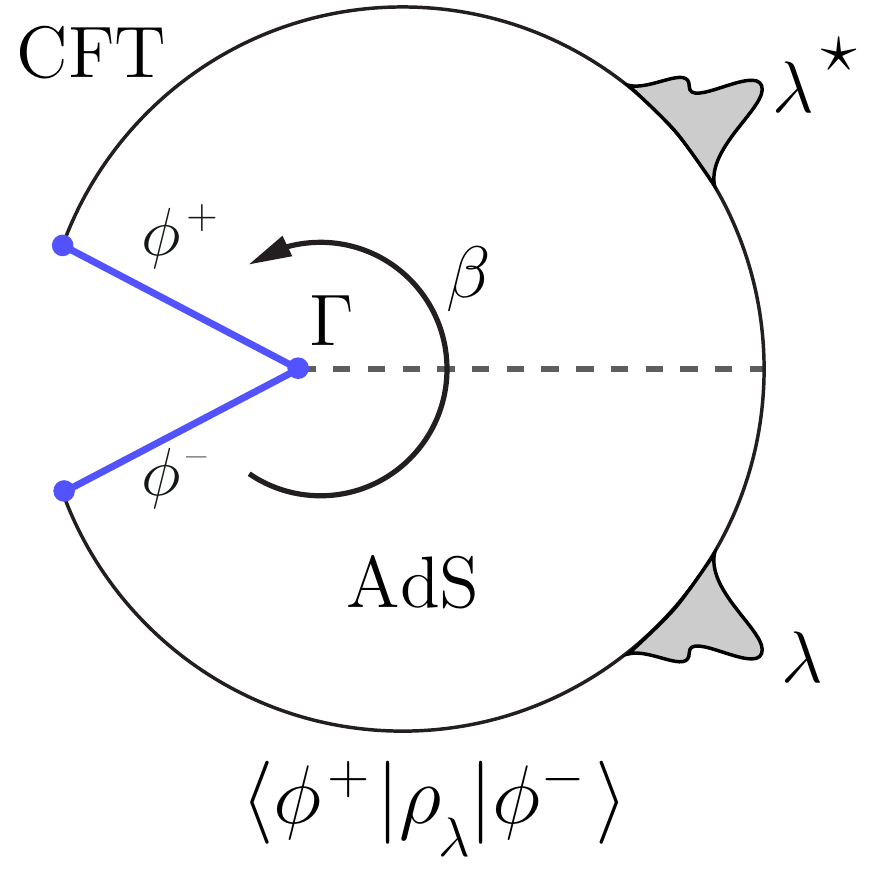}
\caption{}
\end{subfigure}
 \caption{(a) A representation of the state $|\Psi_\lambda\rangle$ as an euclidean path integral is presented. An external source $\lambda\neq0$ prepares an excited state. The CFT subsystem $A$ is denoted in blue, while two candidates for its dual bulk regions are denoted $B_1$ and $B_2$. (b) A depiction of the computation of $\langle \phi^+|\rho_\lambda|\phi^-\rangle$ is shown, where $\rho_\lambda=\text{Tr}_{\bar B}|\Psi_\lambda\rangle\langle\Psi_\lambda|$. The $\phi^{\pm}$ define the field configurations in the branches and $\beta$ is the local angle between them around the codimension-2 surface $\Gamma$ inside the bulk. In building $\langle\Psi_\lambda|$, the source $\lambda^\star(\tau)\equiv\lambda(-\tau)$ must be defined.}
\label{fig:packman}
\end{figure*}

\section{States in holography and decomposition of bi-partite systems}

Consider a local quantum field theory defined on a globally hyperbolic spacetime ${\cal M}= \mathbb{R}_t \times \partial \Sigma$, in a pure state defined through its density matrix $\rho= |\Psi\rangle\langle \Psi|$ we can define the reduced density matrix, $\rho_{A}$, on a subsystem ${A}\in \partial \Sigma$ as the partial trace on the complement of $A$ (denoted by $\bar A$). By definition this object is semi-definite positive and Hermitian and then can be always written as
\begin{equation}
\rho_{A}=\text{Tr}_{\bar A}\rho=\frac{e^{-K_{A}}}{\text{Tr} e^{-K_{A}}}~~,~~~~~\text{Tr}_A \rho_A = 1\;,\label{defirhoA}
\end{equation}
where $K_{A}$ is the modular Hamiltonian.
We will assume that this theory is holographic, i.e, ${\cal M}$ stands for the boundary of spacetimes with fixed asymptotics. Typically one consider the dualty AdS/CFT, where the space $\partial \Sigma$ is compact (a $d-1$-sphere), and the bulk spacetime is asymptotically AdS.

Let us denote as $\Sigma$ the constant-$t$ spacelike hypersurface of the bulk spacetime, and let $B \subset \Sigma$ a \emph{candidate} to the gravity dual of the region $A$ (Fig \ref{fig:packman}b) . The intersection of $B$ with its complement $\bar B \subset \Sigma$, is the codimension-2 (entangling) surface $\Gamma$ that intersects the asymptotic boundary on  $\partial A$. The causal development of $B$ is often called the entanglement wedge.

The (von Neumann) entanglement entropy is computed from \eqref{defirhoA} as
\be\label{defiSA}
S(A)= - \text{Tr}\, \rho_A\, \log \rho_A
\ee
similarly, one can compute the entanglement entropy in the theory of gravity $S(B)$ and by virtue of the holographic correspondence, it should coincide with $S(A)$ for \emph{a suitable} choice of $B$. 

Given a state $\Psi_\lambda$, common to both (gauge/gravity) Hilbert spaces \cite{us4}, the reduced density matrix for a subregion of the boundary (gauge) theory $A$ is obtained by taking the trace on the complement ${\bar A}$; and since the dual of $A$ is $B$, one can naively claim that the holographic dual of this operation is $Tr_{\bar B} \,\Psi \Psi^\dagger$. Nevertheless,
 there is no a clear prescription (at quantum level) on which is the gravitational subsystem $B$ that correspond to the subsystem $A$ on the boundary \footnote{A similar discussion can be found in Ref. \cite{Dong18} to explore the fixed area states subspaces.}.  In a path integral approach, the natural prescription is that one should sum over all the possible partitions of the dual space in two subsystems $B$ and ${\bar B}$ (intersecting in the surface $\Gamma$), such that $B$ intersects the asymptotic boundary on $A$, i.e, to sum over the entangling surfaces $\Gamma$ (see Fig \ref{fig:packman}a).

On the other hand, let us observe that the matrix elements of $\rho(A)$ can be computed in a configuration basis of fields $|\phi\rangle\equiv |\{\phi(x) , \forall x \in B\subset \Sigma\}\rangle$ in the corresponding entanglement subregion $B\subset \Sigma $ of the bulk. This matrix $\rho(B) \equiv\langle \phi^+ |\rho |\phi^- \rangle$ (see Fig \ref{fig:packman}b) can be interpreted as a \emph{representation} of the density operator on ${\cal H}_B$, then, if one changes the subset $B$, the representation is changing\footnote{This is particularly clear in a finite-dimensional Hilbert space (which can be formulated using a suitable discretization of $\Sigma$), where the dimension of the representation would be $d_\Gamma \equiv dim {\cal H}_B$. A construction using OQEC techniques can be found in \cite{Akers18}.}. Thus clearly, these representations can be labeled by the codimension-2 surfaces $\Gamma$.

 Therefore, our prescription here is that the density matrix
of the system $A$, living on the boundary of $\Sigma$, has the structure of a sum over blocks over the different
representations in the bulk
\be \label{rhoA-osum} \rho_\lambda (A) = \bigoplus_{\Gamma}\; \rho_\lambda (B) \;. \ee
In fact one of the results of this work is that the probabilities of the different representations/blocks depend on the area of $\Gamma$ as $e^{-T Area(\Gamma)}$ where $T$ is a real positive number. This resembles the von Neuman's theorem (see e.g. Appendix of \cite{HarlowReview}). Since the algebra of operators in the QFT defined on the boundary is a von Neumann algebra, then in the context of the gauge/gravity duality, it is natural to decompose the Hilbert spaces as 
 \be {\cal H}_A\otimes_{} {\cal H}_{\bar A} \equiv \,\bigoplus_\Gamma \; {\cal H}_B\otimes {\cal H}_{\bar B}\,. \ee
 The objective of this paper is not to study more details of this structure, although interesting questions remain for future research. For the most of applications studied in this work, we are interested in the formula to compute the partition function (and the entropy) in the field theory in terms of the theory of gravity, namely
\be\label{rhoA-dual-prescription} Z_\lambda(A) = \int_{\partial \Gamma = \partial A} [D\Gamma] \; Z_\lambda(B)\; ~~~~~~~~~~~~~~\partial B \equiv\, \Gamma\; \cup \;A \;\;, \ee
which follows from eq. \eqref{rhoA-osum}   by taking trace, on the right hand side one shall sum over the $\Gamma$-blocks. 
This formula expresses that given the subsystem $A$, the surface $\Gamma$ (anchored by $\partial A$) is undetermined \emph{a priori}, and one should sum over all possibilities.

These prescriptions will be useful to relate the entanglement entropies (and modular Hamiltonian) computed to both sides of the gauge/gravity correspondence, and we shall return to them later.

\begin{figure*}
\begin{subfigure}{0.49\textwidth}\centering
\includegraphics[width=.9\linewidth] {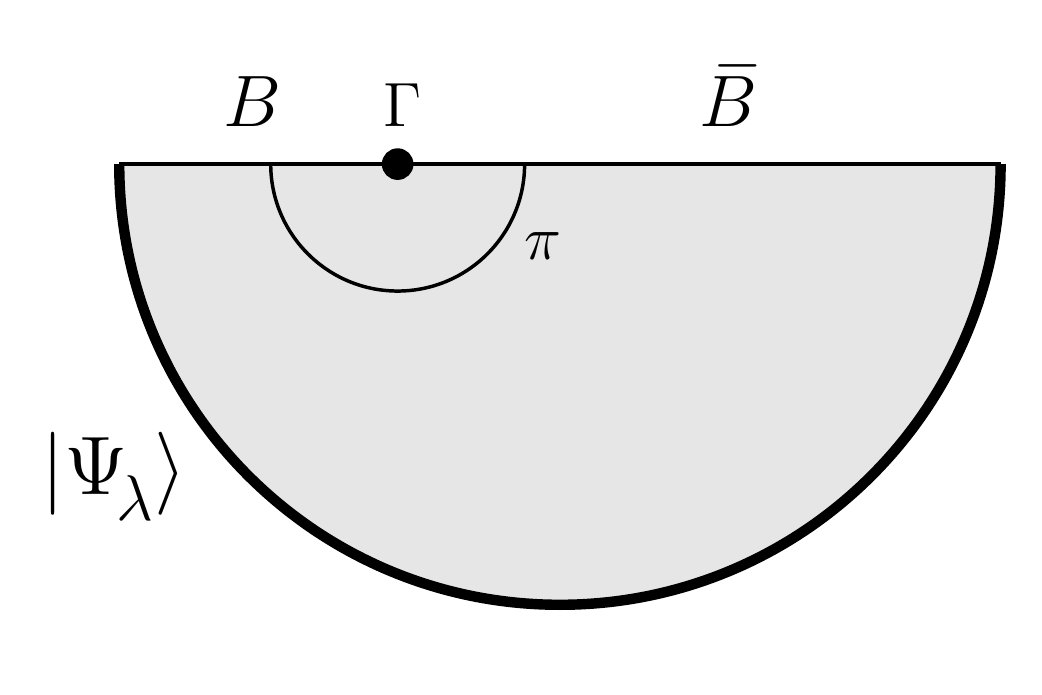}
\caption{}
\end{subfigure}
\begin{subfigure}{0.49\textwidth}\centering
\includegraphics[width=.9\linewidth] {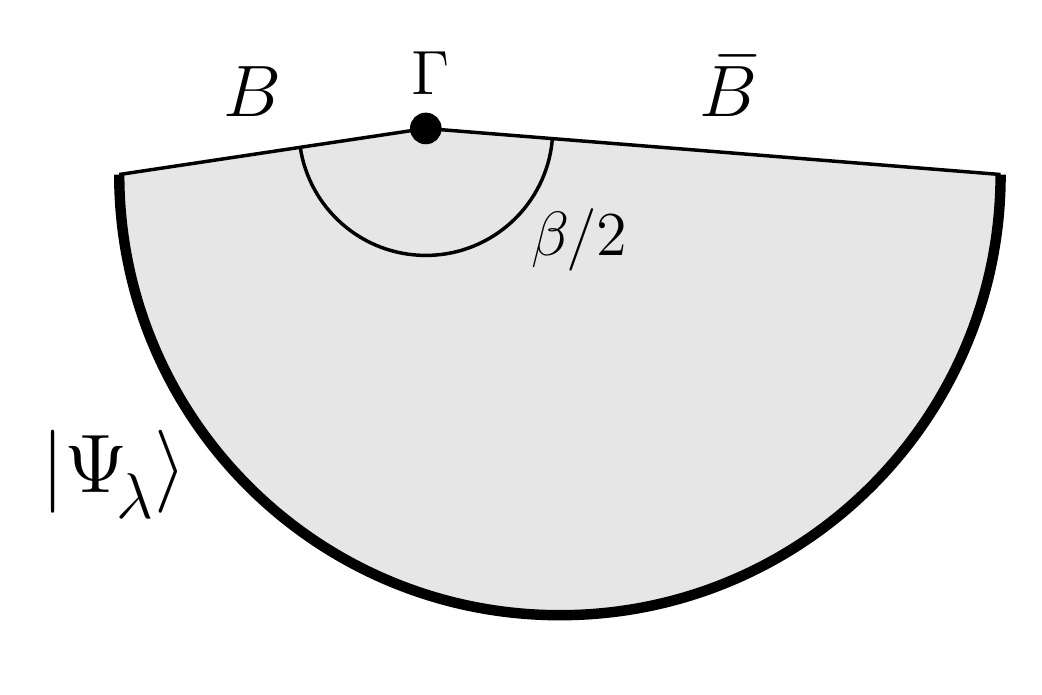}
\caption{}
\end{subfigure}
\caption{(a) The figure shows the state $|\Psi_\lambda\rangle$ projected on a basis of field-configurations defined on a smooth (initial) surface $\Sigma$; while  Fig (b) depicts the projection on a basis $|\phi_\Sigma, \beta\rangle$ associated to a non-smooth surface $\Sigma$, in this case the (Euclidean) path integral that represents the wave functional \emph{requires} a Hayward term in the action.}
\label{fig:H1}
\end{figure*}

\section{Wave functionals in gravity and the Hayward term}

In the present work we will consider states in the field theory whose wave functional can be described as an euclidean path integral in the gravity side such as in the Hartle-Hawking formalism, but with arbitrary  (asymptotic) boundary conditions $\lambda \neq 0$, that correspond to sources on the euclidean extension of ${\cal M}$ \cite{us1, Skenderis}:
\begin{equation}\label{HH-0}
 |\Psi_\lambda\rangle \equiv {\cal P} \,\{e^{-\int_{\tau<0} d\tau \; {\cal O}(\tau) \cdot \,\lambda(\tau)}\}\,|0\rangle  \,\qquad\,\Longleftrightarrow\,\qquad\, \langle\phi_{\Sigma}|\Psi_\lambda\rangle\equiv\int_{(\phi_{\Sigma};\lambda)}{\cal D}\phi \;\; e^{-I[\phi]}
\end{equation}
where $\tau$ denotes the Wick rotated time coordinate of ${\cal M}$. This is the state in the field theory, and the expression on the right is its wave function in the holographic dual. The ground state corresponds to setting $\lambda=0$. These states were extensively studied in different holographic setups \cite{Marolf, us2, Belin, us5}, and extended to finite temperature cases \cite{us3, us4,Mark}.

The path integral on the right is the projection of a state $|\Psi_\lambda\rangle$ onto a of a basis of field configurations on a given
initial spacial surface $\Sigma$. It implicitly supposes the sum over all the euclidean bulk topologies $M^-$ whose boundary is $\Sigma$ and the past ($\tau<0$) of the asymptotic boundary , see \ref{fig:packman}; but at large $N$, only the classical configurations contribute and one evaluates it on
the dominant solution $(M^- , g_{\mu\nu})$.

 The variable $\phi$ here denotes the collection of bulk local fields, including metric and matter fields:
$\phi = (g_{\mu \nu} , \varphi , \dots)$; and $\phi_\Sigma =(h_{ab}, \varphi_\Sigma , \dots)\;, \;\,\lambda=(\lambda_{ab} , \lambda, \dots)$ denote Dirichlet boundary conditions on $\Sigma$ and the asymptotic boundary respectively. For $\lambda\equiv 0$ this describes the Hartle Hawking wave functional for the fundamental state, but it generalizes to other (excited) states as $\lambda\neq 0$ \cite{ SvRL}, which in the large N approximation, correspond to quantum coherent states \cite{us1,us3,us4}.
The total action is 
$I \equiv I_G[g_{\mu \nu} ] + I_{matter} $ where $I_G$ is the gravity action and $I_{matter}$ denote the terms depending on $\varphi$ that would contribute to the action as $o(1/N^k)\;, \, k \geq 0$.

The initial surface $\Sigma$ where one projects the state is arbitrary. The standard choice is a connected and
\emph{smooth} hypersurace, but for our purposes here, will be crucial to consider an initial surface
$\Sigma = B \cup {\bar B}$,
with an angle $\beta/2$ (on $\Gamma$) between $B$ and its complement (see Fig \ref{fig:H1}b).

Let us consider states \eqref{HH-0} such that $M^-$ can be continuously foliated in surfaces $B(\tau)$ 
labeled by an angular parameter $\tau \in [0,  -i\beta/2 ]$ \cite{eternal, us3}, identifying $B(0) \equiv  B $ and $B(i\beta/2) \equiv {\bar B}$. If $\beta/2 \equiv \pi$, the initial surface $\Sigma = B \cup {\bar B}$ is smooth (Fig \ref{fig:H1}a). This geometry is the same that construction Ref. \cite{Takayanagi19} starts with.

The most well known examples of this are: the thermal vacuum \cite{eternal}, and excited (coherent) thermal states \cite{us3,us4}, where the euclidean spacetime can be described by $M^- = B \times [0,  -\pi ]$ and $\tau$ parameterizes a symmetry such that the foliation is uniform: $B(\tau) = B \,,\, \forall \tau$. 
One might alternatively project this state in a basis of configurations of the fields on \emph{another} initial hypersurface $\Sigma' = B \cup {\bar B}$ with angle $\beta/2 \neq \pi$ between $B$ and ${\bar B}$,
but in this case the foliation of the spacetime bounded by $\Sigma'$ cannot be uniform, see Fig \ref{fig:H1}b.

In this context the wave functional can be expressed as a matrix element of an euclidean evolution operator \cite{us3,us4, us5}
\be\label{state-U} \langle\phi_{\Sigma}, \beta/2\, |\Psi_\lambda\rangle = \langle\phi_{B}|\otimes \langle\phi_{\bar B}|\Psi_\lambda\rangle = \int_{\phi_{B} , \phi_{\bar B} , \lambda} [{\cal D} \Phi]_{}\;\; e^{-I[\Phi]} \equiv \langle \phi_{B}| U_\lambda(0, -i\beta/2)| \phi_{\bar B}\rangle \ee
where $\phi_{B}\,,\,\phi_{\bar B}$ are the boundary conditions on $\tau=0$ and $\tau=-i\beta/2$ respectively, then $\langle\phi_{B}|\langle \phi_{\bar B}|$ denotes an element of the (complete) configuration basis of the Hilbert space ${\cal H}_B\otimes {\cal H}_{\bar B}$. The evolution operator shall be seen as a linear map $ U:{\cal H}^{}_B \to {\cal H}_{\bar B}\,$, and according to the gluing rules \cite{us3}, the reduced density matrix can be expressed as the composition at the moment of time-reflection symmetry : \be\label{rho-U} \rho_\lambda \left(: {\cal H}_B \to {\cal H}_B \right) = U_\lambda (i\beta/2 , 0) U_\lambda^\dagger (i\beta/2 , 0)\equiv U_\lambda (i\beta/2 , 0) U_{\lambda^*} ( 0, -i\beta/2) = U_\lambda(i\beta/2, -i\beta/2) \ee where $\lambda^* (\tau) \equiv \lambda(-\tau) $ (see Refs \cite{us1,us3,us4, us5}). For this reason the operator $U$ is also referred to as $\rho^{1/2}$ in the TFD literature \cite{TFD}. The representation of the pure states in terms of evolution operators is convenient and more illuminating for the computations involving the Replica method.

In this case, the full gravity action is expressed as
\be
  \label{IG}
  I_G = -  \frac{1}{16 \pi G_N} \int_{M^-}  \sqrt{g} (R - 2 \Lambda) - \frac{1}{8 \pi G_N} \int_{B} \sqrt{h} K -
   \frac{1}{8 \pi G_N} \int_{\bar B} \sqrt{h} K + \frac{1}{8 \pi G_N} \int_{\Gamma}( \beta/2 - \pi) \sqrt{\gamma}.
  \ee
   where K is the trace of the extrinsic curvature; $\beta/2$ is the angle between the two surfaces $B$ and $ {\bar B} $. 
      The Einstein-Hilbert action with matter will be referred to as the bulk action 
   \be\label{Ibulk} I_{bulk} [\phi , M^-]\equiv \frac{1}{16 \pi G_N} \int_{M^-}  \sqrt{g} (R - 2 \Lambda)  + I_{matter} [g, \varphi, \dots]\ee
   which includes all the integrals on the points of the \emph{interior} of $M^-$. The boundary contributions are given only by the Gibbons-Hawking term and extra (local) contributions of matter fields on the boundaries of $M^-$
   \be\label{Ibdy} I_{bdy} [\phi_\Sigma , \lambda, \partial M^-] \equiv  \frac{1}{8 \pi G_N} \int_{\bar B} \sqrt{h} K + \frac{1}{8 \pi G_N} \int_{B} \sqrt{h} K + I_{matter} [h, \varphi_\Sigma , \lambda, \dots] ,\ee
and the so-called Hayward term (see refs \cite{Takayanagi19, Hayward93}):
\be 
\label{2.4}
 I_H (\Gamma) =  \frac{1}{8 \pi G_N} \int_{\Gamma}(\beta/2 - \pi) \sqrt{\gamma}\,\;\;,
 \ee
 which vanishes for $\beta/2 \equiv \pi$ that describes a smooth (without wedges) initial surface $\Sigma$. 
In this first study we will set to zero the gravitational sources at the asymptotic boundary ($\lambda_{ab} \equiv 0$) for simplicity, and so the asymptotic gravitational terms do not appear in  the action \eqref{Ibdy}.
The classical problem for this theory is well posed by fixing Dirichlet boundary conditions on $B$, ${\bar B}$, the asymptotic boundary, and the angle $\beta/2$, and then $\gamma$ on $\Gamma$ is computed from the bulk metric.

Then in the large $N$ limit, the gravitational (unnormalized) wave function can be computed in the saddle point approximation:
\be 
 \Psi_\lambda \left(\phi_B, \phi_{\bar B}, \beta/2 \right) \left( = \langle \phi_{B}| \,U_\lambda\left(-i\beta/2, 0\right)\,  |\phi_{\bar B}\rangle \right)\, = e^{-I_{bdy}[\phi_B, \phi_{\bar B}, \lambda] + \frac{(\pi - \beta/2) }{8\pi G} a(\Gamma_{})} , \label{psi-BBbar-largeN}
\ee
where we have imposed that the opening angle (between $B$ and ${\bar B}$) is uniform along the surface $\Gamma$ and the Hayward term is\footnote{This requirement implies that $\beta$ projects on the intersection of $\Gamma$ with the asymptotic boundary, so it also characterizes configurations basis of the QFT defined on the boundary.}
\be\label{Hay1}
I_H[\Gamma,\beta/2] = -\frac{1}{8\pi G}\int_\Gamma (\pi - \beta/2) \sqrt{\gamma} = -\frac{(\pi - \beta/2)}{8\pi G} a(\Gamma)
\ee
At this point we would like to point out that the Hayward term appears as the state is projected on a particular basis $|\phi_\Sigma , \beta/2  \rangle$ of configurations, thus it appears as a property of the wave functionals of gravity (or components) by projecting the state on a specific  basis, rather than about the state itself.
The state $\Psi_\lambda$ of the dual field theory, is characterized by the the sources $\lambda$ on the interval $(0, \pi)\times \partial \Sigma$ of the asymptotic boundary.

\section{The gravitational density matrix and the \emph{area operator}}

The reduced density matrix associated to the region $B$ of the bulk is 
 \be \rho_\lambda (B, \beta) \equiv \text{Tr}_{{\cal H}_{\bar B}} \,|\Psi_\lambda \rangle \langle \Psi_\lambda |
= \sum_{\phi_{\bar B}} \,  \langle\phi_{\bar B}|\,\Psi_\lambda \rangle\, \,\langle \Psi_\lambda \,| \phi_{\bar B}\rangle \, .\ee
 Defining two arbitrary field configurations $\phi^\pm \equiv \phi(B^\pm) = \phi(\pm i \beta/2)$ on two copies (or branches) of the surface $B$, denoted as $B^\pm$ (Fig 1b), which intersect in a co-dimension two surface $\Gamma = B^+ \cap B^-$, and using the relation \eqref{state-U}, one can express its matrix elements as the product of euclidean evolution operators (eq. \eqref{rho-U}):
\be\label{rho-psipsi} \langle \phi^+ | \rho_\lambda (B, \beta) |\phi^-\rangle = \sum_{\phi_{\bar B}}
\langle \phi^+_{}| \,U_\lambda\left(-i\beta/2, 0\right)\,  |\phi_{\bar B}\rangle\langle\phi_{\bar B}| \,U_\lambda\left(0, i\beta/2)\right)\,| \phi^-_{}\rangle = \langle \phi^+ | U_\lambda\left(-i\beta/2 , i\beta/2 \right) |\phi^-\rangle
\ee
where we have used the completeness of the configuration basis $I_{\bar B}\equiv\int {\cal D}\phi_{\bar B}|\phi_{\bar B}\rangle\langle\phi_{\bar B}|$ on ${\cal H}_{\bar B}$. This is well defined as a path integral, 
and one can compute this in the Large $N$ approximation,
and using \eqref{psi-BBbar-largeN}:
\be\label{rhoB} \langle \phi^+ | \rho_\lambda (B, \beta) |\phi^-\rangle = \int^{}_{\phi^\pm , \lambda}[{\cal D}\Phi] \; e^{-I[\Phi]}
\approx \, e^{-I_{bulk}[\phi, M]} \, e^{-I_{bdy}[\phi^\pm , \lambda] + \frac{(2\pi - \beta) }{8\pi G}a(\Gamma_{})}
\ee
where, by virtue of the saddle point approximation, we evaluated the action $I_G$ in a classical solution $M = M^- \cup M^+$ smoothly glued on the surface ${\bar B}$, whose boundaries are the branches $B^-$ and $B^+$ (see Fig \ref{fig:packman}a). 
Obviously the boundary data $\phi^\pm$ that label the matrix elements (so as $\lambda$ characterizing the state), backreact with the bulk metric.
Notice that because of \eqref{psi-BBbar-largeN} and \eqref{rho-psipsi}, in the exponent of this expression  appears as a sum of two (equal) Hayward terms \cite{Takayanagi19}.

Let us see briefly that in the present formalism $a(\Gamma)$ shall be interpreted as an operator. In fact, the Hayward term in the action \eqref{IG} and in the wave functional is crucial to it.

The opening angle $\beta/2$ (actually, its analytical extension $-i \beta/2$) and the volume element $\sqrt{\gamma}$ can be taken as the variables canonically conjugated in the ADM formalism, associated to ``edge'' modes (see Ref \cite{Takayanagi19} for details); then, in an eventual canonical quantization of gravity as these quantities be promoted to operators \cite{ensayoarea, Jafferis}:
\be\label{A-on-psi} \langle \phi_\Sigma , \beta| a(\Gamma) | \Psi\rangle =  -(8\pi G )\, 2\frac{\partial }{\partial \beta}\Psi(\phi_\Sigma , \beta)\ee
We see that by variate the wave functional eq. \eqref{state-U} with respect to $\beta$  one obtains the action of the area operator on the global state. Then obviously the computation of it depends on which $o(G)$ approximation the path integral \eqref{state-U} is being calculated. For instance, to leading order the wave functional is given by the rhs of \eqref{psi-BBbar-largeN} and the area of $\Gamma$ is nothing but the area computed with the induced metric $\gamma$, obtained from the boundary data: $h_B, h_{\bar B}$ by continuity.  This shows that in the present set up $a(\Gamma)$ can be considered an operator, and many calculations are precisely defined, as the Hartle-Hawking path integral can be better calculated. For example, its expectation value can be computed from eq. \eqref{rhoB}  by taking the trace (summing over the gluing conditions $\phi^+=\phi^-$), and differentiate it with respect to $\beta$:

\be\label{expectation-Aope}  \text{Tr}\{\, \rho(B,\beta) \, a(\Gamma)\}  =  -(8\pi G ) \frac{\partial }{\partial \beta} \text{Tr}\, \rho(B, \beta) = - (8\pi G ) \frac{\partial }{\partial \beta} Z(B, \beta)\;.\ee
This can be considered an explicit realization of previous proposals viewing the area as an operator \cite{ensayoarea,Jafferis,JLMS,Dong18}.

\section{The partition function and gravitational entropies}

Expression \eqref{psi-BBbar-largeN} is the reduced density matrix associated to the entanglement region $B$ -with boundary $\Gamma$-, and clearly the angle between the boundaries $B^-$ and $B^+$ is $\beta$. 
To evaluate the partition function $Z(\beta)\equiv \text{Tr} \rho$ in the large N approximation, $B^+$ and $B^-$ must be smoothly glued, such that their contributions (the Gibbons-Hawking terms) cancel out \cite{SvRL, us1, us2}.

It is worth pointing out that in the standard previous computations of this partition function (as function of $\beta$), the Hayward term is ignored,
 and one would obtain a conical geometry with only one asymptotic boundary, and a deficit angle $2\pi - \beta$. Thus the total action is \eqref{Ibulk} and only the tip of the cone contributes to the action with a scalar curvature $R= 4\pi \, (1 - \beta/2\pi ) \, \delta_\Gamma$, so the on-shell action is proportional to the area of $\Gamma$ \cite{MaldaAitor,Takayanagi19}, such that 
 \be\label{ZB} Z= Z_{bulk}\; e^{ \frac{a_\Gamma\, }{8\pi G}(2\pi-\beta)}\; .\ee
If one consider the vacuum state $\lambda\equiv 0$, then $\phi=0$ everywhere, and \,$\log Z_{bulk}$\, is given only by the gravity the action that goes over the regular part of $M$. The contribution of the cosmological term can be eliminated by normalizing the state $\rho \to \rho/ Z(1)^{}$ (see Sec 6).

So, the gravitational (von Neumann) entanglement entropy in this case is independent on the range $\beta$
\be\label{S-vN-B}
S(B) = \log Z -\beta \frac{\partial \,\log Z}{\partial \beta}= \frac{a_\Gamma\, }{4\pi G}
\ee
which is the expected area law, and the derivation is similar to \cite{Fursaev}. 
A criticism with this is that 
on-shell contributions to the path integral coming from conical geometries (with $\beta\neq 2\pi$), should require suitable sources in the bulk that cannot be justified only from the  theory of gravity \eqref{IG} \cite{Headrick,MaldaAitor}\footnote{In a \emph{pure}-gravity path integral, the dominant contributions are smooth (vacuum) solutions.}. Then one need to consider some appropriate extension of the theory to include back reacting fields such  that effectively behaves as a cosmic brane  \cite{Dong16}, although it is difficult to argue that using the standard holographic recipes.

In contrast in the present formulation, we have just shown that the construction of a state $\rho(B)$ where the interval $\beta$ differs from $2\pi$, \emph{requires} a Hayward term. The total theory of gravity that one shall consider is $I_{bulk} + I_H(\Gamma)$, such that $\Gamma$ is taken as a dynamical variable (as argued in the next Section), and so there are \emph{classical solutions} with conical singularities, avoiding the criticism mentioned above.

In this case, the result \eqref{ZB} is recovered as follows. The surfaces $B^\pm$ are identified after a period $\beta$, so $M$ is a conical geometry but $\Gamma$ is part of the boundary. The Einstein-Hilbert (bulk) term is local and integrates the scalar curvature in the \emph{interior} of $M$, where it is regular. 

Thus, since the total gravity action is \eqref{Ibulk} (without the Gibbons-Hawking terms) results that
\be\label{ZB-withH} Z (B, \lambda, \beta) \equiv \text{Tr}_B\, U_\lambda (i\beta) = Z_{bulk}[M]\; e^{ \frac{\, (2\pi-\beta)}{8\pi G}a_\Gamma}\; ,\ee 
is the partition function associated to a region $B$ of the bulk with fixed boundary $\Gamma$. But we will show below (using the replica method) that the cosmological term does not contribute, and the other contributions to the prefactor $Z_{bulk}[M] $ can be neglected such that only the Hayward term is relevant for the computation \eqref{S-vN-B}. These results will be recovered in Sec 6 using replicas, and many details will be clarified.

\pagebreak
\textbf{Density matrix and entropy in the boundary QFT}
\vspace{0.5cm}

The previous results in the gravity side suppose an arbitrary (fixed) separation in two subsystems of the bulk d.o.f's : $B\cup{\bar B}$ and a entangling surface $\Gamma$,
but now we shall translate them to the field theory defined on the boundary. As argued in Sec 2,
the \emph{gauge/gravity} duality prescribes that the respective Hilbert spaces are equal; therefore, we are implicitly assuming that the state is described by the same object $|\Psi_\lambda\rangle$ \cite{us1, us5}. However, this state can have very different representations in the \emph{gauge} or \emph{gravity} theories, and we actually do not know precisely how they relate. A particularly relevant issue about this is how to relate the reduced states on the regions $A$ and $B$ and the respective partition functions, although in Sec 2 we argued a formula for it, involving a \emph{sum} over $\Gamma$'s (eq \eqref{rhoA-dual-prescription}).

The main result of this work is that the Hayward term, added to the prescription \eqref{rhoA-dual-prescription}, effectively explains the cosmic brane proposal of \cite{Dong16}, and accounts for the entanglement entropies in the field theory defined on the boundary. Moreover, the proposal \eqref{rhoA-osum} is crucial to define the boundary density matrix, such that the area appears as an operator \cite{Jafferis,JLMS}.

In fact,
plugging \eqref{rhoB} in the prescription \eqref{rhoA-osum}
we obtain the (path integral) formula for the reduced density matrix in the boundary:
\be\label{rhoB-sumGamma}  \rho_\lambda (A) \; = \;\bigoplus_{\Gamma} \;\, \int^{}_{\phi^\pm (\Gamma) , \lambda}[{\cal D}\Phi] \; e^{-I[\Phi]}
\approx \, e^{-I_{bulk} [M]} \, e^{-I_{bdy}[\phi^\pm , \lambda] + \frac{(2\pi - \beta) }{8\pi G}a(\Gamma_{min})} \;\oplus\; \dots  
\ee
where the right hand side is the large $N$ approximation of the most probable representation $\Gamma\,(\partial \Gamma = \partial A)$, associated to the surface of minimal area. So the field configurations $\phi^\pm \equiv \phi^\pm (\Gamma_{min}) $ refers to the matrix elements in that representation (and "$\dots$" denote the others less probable).
The prescription \eqref{rhoA-dual-prescription} consists of taking the trace of this, and the result is the sum on the $\Gamma$-blocks of the traces of each sector $\text{Tr}_B \rho_\lambda(B) $, that are computed summing over $\phi^+(B)=\phi^-(B)$ (this operation is equivalent to glue the surfaces $B^\pm$ after an interval $\beta$ \ref{fig:packman}). So we obtain the partition function associated to the region $A$ in the boundary theory:
\be\label{Z-A-final}
Z(A, \beta) = \int_{\partial \Gamma = \partial A} [D\Gamma]\int^{}_{ \lambda}[{\cal D}\Phi] \,e^{-I_{bulk} [\Phi]-I_{bdy}[\lambda] - \frac{(2\pi - \beta) }{8\pi G}a(\Gamma_{})} \approx \,e^{-I_{bulk} [M]-I_{bdy}[\lambda] + \frac{(2\pi - \beta) }{8\pi G}a(\Gamma_{min})}
\ee
where, on the r.h.s we have used the saddle point approximation and evaluated on the surface whose area is a minimum $\Gamma_{min}$ (for $\beta > 2\pi$). Finally, we compute the entanglement entropy for the region $A$ using the formula \eqref{S-vN-B} and  obtain the RT formula 
\be
S(A)= \frac{1}{4 G}a(\Gamma_{min})\;\;.
\ee

The computation of \eqref{rhoB-sumGamma} and \eqref{Z-A-final} is well defined since the boundary problem indicated in Fig. \ref{fig:packman} is well posed. Notice that the problem consists in solving the coupled system of equations for the fundamental fields $\Gamma, g, \phi$, derived from a Nambu-Goto action coupled with gravity with a tension 
\begin{equation}\label{tension}
T = \frac{2\pi-\beta}{8 \pi G} \;\;.
\end{equation}
such as in the formulation \cite{Dong16}. The difference is that the Hayward term replaces the cosmic brane, but it is part of the gravitational action
rather than an artifice to find the classical geometry with the suitable conical singularity. In contrast with the approach of \cite{Dong16}, this term  
contributes crucially to the partition function and to the direct computation of the Rényi entropies, presented in Sec. 6. 

Remarkably, the formula \eqref{tension} for the real-valued tension as function of the opening angle $\beta$ between the branches $B^+$ and $B^-$ is universal, and it open the possibility of interesting generalizations (e.g. higher order gravity); in particular we will see below that it works as a parameter to generalize the von Neumann gravitational entropies. In the limit as $\beta \to 2\pi$, the geometry of Fig \ref{fig:packman}(b) is given by a solution of gravity, and the cosmic brane becomes the non-backreacting minimal surface of the RT prescription. The presence of the Newton constant in the brane tension suggests that the nature of this term is \emph{gravitational}, which enforces our point of view.

\section{Rényi entropies from Hayward term using replicas}

In this section we use a version of the replica method
to compute the spectrum of Rényi (and von Neumann in the limit $n\to 1$) entropies using the Hayward term. It is markedly different from the calculus \cite{Takayanagi19}, and closer to the method of refs \cite{MaldaAitor,Dong16} but where the Hayward term plays a crucial role.

Let us consider now the euclidean spacetime solutions $M_n$ of $n$ copies of the asymptotic  boundary conditions: $\lambda_n\;$ on $\partial M_n \equiv (0, 2\pi) \,\cup \,(2\pi , 4\pi)\, \cup \,\dots (2\pi(n-1) , 2n\pi)\;\times  (\partial \Sigma)_{(d)} $, where $0$, and $2n\pi$ are identified; i.e: th BC's of a single copy, $\lambda$, is repeated $n$ times before gluing the boundaries $B^\pm$. To identify the edges $B^\pm$ corresponds to take the trace of \eqref{ReplicaT1}, and the geometry $M_n$ becomes periodic (with period $2\pi n$).

Since the un-normalized matrix density $\rho^n$ can be described as the evolution operator \cite{MaldaAitor, us3, us4}, at large $N$, the bulk computation consists in evaluating the (euclidean) path integral on the classical solution:
\be \label{ReplicaT1}
\text{Tr}\; \rho^n (B) = \text{Tr}\; U(B, i (\beta = 2n\pi)) = Z[B, M_n] \approx e^{-I[M_n]}  \;.
\ee
In our mindset (e.g. eq \eqref{rho-psipsi}), this expression can be thought as a bra-ket of a initial state with $\beta/2 \equiv n\pi$; thus, the semi-classical approximation \eqref{rhoB} has a Hayward term proportional to $\delta_n = 2\pi(1 - n)$.
So therefore, the classical dominant solution $M_n$ is a conifold with deficit angle $\delta_n$ (and tension $\delta_n / 8\pi G$) and the repeated boundary conditions $\lambda_n$.

One can directly observe that the logarithm of the left hand side of eq. \eqref{ReplicaT1} is proportional to the $n^{th}$ order R\'enyi entropy. Nevertheless, the right computation involves the normalized density matrix, and requires to divide this expression by the number $(\text{Tr} \,\rho )^n$.
The normalized density matrix is \be\label{rhotilde}\tilde{\rho} \equiv \rho/ Z(M_1)^{}\,,\ee where
\be\label{norm1} -\log \, \text{Tr} \, \rho(M_1) \equiv -\log Z(M_1) = I[M_1]  \ee and $I$ is given by \eqref{IG}. Then using \eqref{ReplicaT1} we have
\be \label{norm2} \log \text{Tr}\, \tilde{\rho}^n = \log\, Z[M_n] - n \log \,Z(M_1) = I_{bulk} [M_n] - n I_{bulk} [M_1] + I_H(M_n) - nI_H(M_1)\ee
but noticing that for $n=1$, $2\pi - \beta = 0$, the contribution from the Hayward term  is the same upon normalization, i.e 
\be\label{IH-normal}\tilde{I}_H \equiv I_H(M_n) - nI_H(M_1) =I_H(M_n) = 2\pi (1-n) \, a(\Gamma_{min, n})\;\;.\ee
Note that since we aim at the field theory partition function, (according to eq. \eqref{Z-A-final}) the Hayward term here is valued on-shell on the minimal surface $\Gamma_{min, n}$ in the target spacetime $M_n$. If one is interested in the purely gravitational computation $Z(B)$, it shall be valued on arbitrary $\Gamma$ (see \cite{Dong18}).
The bulk terms are
\be
    \tilde{I}_{bulk}[M_n] \equiv I_{bulk}[M_n] - n I_{bulk} [M_1]  = \left(  I_G (g_{(n)}, M_n)\, -  \, n \, I_G (g_1 , M_1)\right) + \tilde{I}_{Matter} [\phi_{(n)}, M_n]\,\,.
\ee
where 
\be \label{IGonshellMn}
\tilde{I}_{G}[M_n] =  \left( I_G (g_{(n)}, M_n)\, -  n\,I_G (g_1 , M_1) \right) = \frac{1}{16\pi G}\left(\int_{M_n} (R_n-2\Lambda) \sqrt{g_{(n)}}\, - \, n \, \frac{1}{16\pi G}\int_{M} (R-2\Lambda) \sqrt{g_{(1)}}\right)\,\,.
\ee

There is not  asymptotic boundary terms in this expression, because the asymptotic boundary condition for $M_n$  has been defined as $n$- equal copies of $\lambda_1$ (on $\partial M_1$). In Einstein gravity ($\Lambda=0$), \eqref{IGonshellMn} vanish trivially on a \emph{vacuum} solution ($\varphi =0$), and then \eqref{norm2} is given only by the Hayward term.
 For gravity with cosmological constant, this action is proportional to the volume and the volume of $M_n$ is proportional to $n$, then we also have that the combination $\Lambda ( Vol[M_n] - n \, Vol[M_1]) $ vanishes 
\footnote{This can easily understood by considering the metric $g_n$ of $M_n$ in the region near the tip: $ds^2 = dr^2 + r^2 d\tau^2 + \gamma_{ij}dx^i dx^j $, $0\leq \tau \leq 2\pi n$, which is locally independent on $n$, so the total volume is: $Vol[M_n] = n Vol[M_1]$. Away from the singularity this relation is trivial because the solutions $M_n$ are simply copies of $M_1$.}.
Consequently, the calculation \eqref{norm2} finally results 
\be \label{norm-final-n} \log\, \tilde{Z}[M_n] = \log \text{Tr}\, \tilde{\rho}^n = \, I_H(M_n) + \tilde{I}_{Matter}\;,\ee that can be analytically extended 
 to real values $n\to \beta/2\pi$, as discussed around eq. \eqref{ZB-withH} (see refs \cite{Dong16,Dong18}):
\be \label{norm-final-vacuum} \log\, \tilde{Z}[M_{\frac{\beta}{2\pi}}] =  (2\pi -\beta) \frac{a(\Gamma_{})}{8\pi G} + o(G^{k\geq 0})\,.\ee
We see that the only contribution to the leading order ($1/G$) of the normalized partition function is given by the Hayward term.
 Even if $\varphi \neq 0$ (with non trivial asymptotics: $\lambda \neq 0$) one can ignore the back-reaction since it contributes to subleading terms ($\sim o(G^{k\geq 0})$).

\subsection{Calculus of the Rényi entropies}

The main result of this Section can be directly observed from expression \eqref{norm-final-n}, and \eqref{IH-normal}. Using that the contribution to the (normalized) partition function of the term $\tilde{I}_{bulk}$ is neglected and only the Hayward term contributes:
\be -\log \,\text{Tr} \tilde{\rho}^n = \delta_n \, \frac{a(\Gamma_{n , \,min})}{8\pi G} \ee
where $\delta_n = 2\pi (n-1) $, then from  the definition of the standard n$^{th}$ order Rényi entropy \eqref{defirenyi} one obtains
\be\label{renyi-result}
S_n = \frac{1}{1-n} \,\log \, \text{Tr} \tilde{\rho}^n = \frac{a(\Gamma_{n\,,\, min})}{4G}\;\;,
\ee
which is a (minimal) area law in agreement with the conjecture of  Ref. \cite{Myers}, where the minimal surface backreacts with the geometry but in this prescription, remarkably, the space time $M_n$ is the solution to the $n$-replicated boundary condition and the tension localized on the surface  $\Gamma_{}$ is
\be\label{tension-renyi}
T_n = (n-1)/ 4G\;\;,
\ee
acting as a gravitational source \cite{Vilenkin}.
The Ryu-Takayanagi prescription is recovered from this computation in the limit $n\to 1$: The tension vanishes in this limit, and the minimal surface $\Gamma_{n\,,\, min}$ reduces to  $\Gamma_{ min}$, which is the minimal (non-backreacting) surface in the smooth space time $M_1 = M$ (without conical singularity).

There is a particular class of geometries such that this computation gives an \emph{flat} spectrum, since the area of $\Gamma_{n\,,\, min}$ is independent on $n$, and our formula agrees with results of \cite{Dong18} (see example in Appendix: eq (A2)). 
In this case the modified Renyi entropy computed from \eqref{defirenyi-hat}, also coincides with \eqref{renyi-result}. This is related to the fact that the quotient of the spacetime $M_n /Z_n $ coincides with $M_1$.

Finally, we note that for a fixed region $B \subset \Sigma$, one is computing the spectrum of gravitational Renyi entropies associated to this subset, thus \eqref{renyi-result} is given by the (fixed) area of the surface $\Gamma\equiv \partial B \cap M_n$

\subsection{The prescription for the modified Rényi entropies}

The definition of the modified n$^{th}$-Rényi entropies eq. \eqref{defirenyi}, can be conveniently put in a more familiar form
\be\label{S-hat-n}
\hat{S}_n = \log Z  - n \frac{\partial \,\log Z }{\partial n} = \left. \left( \log Z  -\beta \frac{\partial \,\log Z }{\partial \beta}\,\right) \;\right|_{\beta = 2\pi n}
\ee
which shows it  explicitly as a one-parametric extension of the von Neumann entropy. This expression can be analytically extended
to any real value $\beta \geq 0$ (coinciding with the extension of the von Neumann entropy and the thermodynamic picture \cite{Dong16,Renyi}), and the discrete spectrum $\hat{S}_n$ is recovered by taking $\beta \equiv 2\pi n$ at the end of the calculation.
Moreover, notice that in the present approach, the geometric interpretation of the analytically extended parameter $\beta$ is the period around the conical singularity.  

In this Section we will show that the model of Ref. \cite{Dong16} can be recovered in the present set up, and therefore, the correct prescription for the modified n$^{th}$-Rényi entropies. To achieve this goal we must have into account some subtleties on the gravitational measure and the path integral, for instance, the calculus of $\text{Tr} \, \rho^n$ through replicated dual geometries involves a manifest discrete symmetry $Z_n$. 

In the holographic replicas construction, the boundary is replicated $n$ times and one could permute \emph{cyclically} these copies, obtaining the same boundary condition $\lambda_n$ and the same euclidean spacetime $M_n$ filling it.
This is the meaning of the so-called replica symmetry $Z_n$. 
Therefore, let us consider spacetimes $M_n$ with this symmetry, such that one can define the orbifold:
\be\label{ReplicaT2-vfinal}
\hat{M}_n \equiv \frac{M_n}{Z_n} \;.
\ee 
In fact, the geometry $M_n$ quotiented by the replica symmetry $Z_n$, satisfies the un-replicated boundary conditions associated to the original spacetime: $\lambda_1$ on $\partial M_1 = [0, 2\pi] \times \partial \Sigma $, but it has a conical singularity on
the codimension-2 transverse surface $\Gamma$, which consists of the fixed points of the replica symmetry \cite{Dong16}.
Moreover, using the locality of the bulk action, we have:
\be\label{cociente-replica}
I_{bulk}[M_n] = n I_{bulk}[\hat{M}_n]\;.
\ee
This property is not extensive to the Hayward term, but one can write $I[M_n] = n I[\hat{M}_n]$ by \emph{defining}
\be\label{I-hat} I_{}[\hat{M}_n, \Gamma] = I_{bulk}[\hat{M}_n]  + \frac{\hat{\delta}}{8 \pi G}\,\, a(\Gamma)  \ee 
where $\hat{\delta} \equiv  2\pi (1-n) /n$. 
For our calculation below, we only need demand that the replica symmetry holds at the level of the action, 
in line with the assumption of previous derivations \cite{MaldaAitor,Dong16}, which assume that the $Z_n$-symmetry is not spontaneously broken by the dominant solution. In other words, the relation \eqref{cociente-replica} is satisfied for the \emph{off-shell} geometries considered in the path integral.

At this point, it is illuminating to write down the partition function. 
Considering only gravity for simplicity, we can express the path integral \eqref{Z-A-final} as
\be\label{Z-hat-A-withX}
Z(A, \lambda_1 , n) = \int_{\partial X|_{\partial M} = \partial A} [DX]\int^{}_{ \lambda}[D M_n] \,e^{- n I_{bulk} [\hat{M}_n ]+\frac{2\pi(1- n)}{8\pi G}a[X, \hat{g}(X)]}
\ee
In this expression $\hat{M}_n$ stands for a spacetime equipped with the corresponding metric $\hat{g}$ obtained from $g$ (of $M_n$) by the quotient \eqref{cociente-replica}. Generally, both metrics are the same locally but the range of the coordinates is different.

The embedding fields $X(\Gamma)$ were put explicitly here in order to highlight which are the fields in the action and differentiate them from the parameters. Recall that the last term  is nothing but a Nambu-Goto action where $a=a[X , \hat{g}_{}(X)] $. Thus this partition function is only a function of the asymptotic boundary conditions $\lambda = \lambda_1$, the subset $A$ of $\partial \hat{M}_n = \partial M_1$ where the QFT lives, and the number of replicas $n$, while the fields $X, \hat{g}_{\mu \nu} , \varphi$ are integrated out.

The difference of this partition function with $Z(A, \lambda_n , n)$ considered previously is that, although it also sums over spacetimes whose boundary is the branched cover $\partial M_n $ (with replicated boundary conditions $\lambda_n$), the replica symmetry is considered manifest and one must sum on geometries as \eqref{cociente-replica}. This guarantees that the replica symmetry is not spontaneously broken \cite{MaldaAitor,Dong16}. Note that 
there is a sort of redundancy in the measure because one the sums over the replicated geometries $M_n$'s, however here we need not more technical details on this.

Let us consider now the saddle point approximation in this context. Factorizing out $n$ in the total action appearing in \eqref{Z-hat-A-withX}, and since $n>0$, the dominant solution is obtained by minimizing \eqref{I-hat}. Then we
have 
\be\label{Ihat-saddle} \frac{1}{n} \log \text{Tr} \rho^n = \left. I_{}[\hat{M}_n, \Gamma] \right|_{on-shell}  + (\dots) \;,\ee
where the rhs is the action valued on a solution of the coupled theory \eqref{I-hat}, and $(\dots)$ denotes (quantum) corrections ($o(G^{k\geq0}$) to the saddle point approximation.  
Notice that the Hayward term must be valued on the minimal surface $\hat{\Gamma}_{min}$, whose area is a minimum \emph{in the dominant} geometry $\hat{M}_n$, with a deficit angle $\hat{\delta}$. This coincides \emph{exactly} with the action of the model in \cite{Dong16}, given by an action of gravity plus a cosmic brane with a tension corresponding to the same deficit angle: $ \hat{\delta} / 8 \pi G $.
Finally, the modified $n^{th}$-Rényi entropy is computed using the formula \eqref{S-hat-n} or $\hat{S}_n = -n^2 \frac{\partial}{\partial n } n^{-1} \log \hat{Z}(A,\lambda, n)$ ( eq. \eqref{defirenyi-hat}).

Noticing that we actually need to take a derivative of the path integral \eqref{Z-hat-A-withX} with respect to the parameter $n$ (or $\beta$)\footnote{Taking $n$ as a parameter, the derivation is off-shell and we only shall take derivatives of the quantities that depend explicitly on $n$ in the total action. Moreover, the bulk gravity action is an integral on the local curvatures and metrics off-shell that are integrated out, then the dependence with $n$ only can be in the limits of integration that, because of the quotients by $Z_n$, are typically independent on $n$ (e.g, $\int_0^{2\pi}d\tau$).}, and using that $I[\hat{M}_n]$ and $a[\Gamma, \hat{g}]$ are (off-shell) independent on $n$, we obtain the expected result 
\be\label{resultado-S-hat} \hat{S}_n =  \left \langle -n^2 \frac{\partial}{\partial n }  I \right \rangle = -n^2 \left \langle \frac{\partial}{\partial n}\frac{\hat{\delta}}{8 \pi G}\,\,  a(\Gamma) \right \rangle  = \frac{1}{4 G}\langle\, a \,\rangle = \frac{a(\hat{M}_n , \Gamma_{min})}{4 G} + o(G^{k\geq 0})\ee
for the vacuum state ($\lambda \equiv 0$), although it is straightforwardly generalizable to excited states \cite{us1,us3}. The bracket $\langle \dots \rangle$ stands for the object \emph{within} the path integral \eqref{Ihat-saddle}, thus on the rhs, the expectation value of the area was approximated 
by its value on the dominant classical solution $\hat{\Gamma}_{min}, \hat{g}$.
It is worth emphasizing here that the manifold associated to this solution (denoted as $M_n$) should not be confused with the saddle of the replicated boundary condition of the prescription \eqref{renyi-result} (Sec. 6.1). In Appendix we give examples in $d=2+1$ where, except for $\hat{\delta}=0$, the solution $\hat{M}_n$ is the quotient of a \emph{smooth} geometry $M_n$ (see also \cite{MaldaAitor, Dong16}).

This result agrees with the formula for $\hat{S}_n$ derived  previously by Dong \cite{Dong16}. The difference is that our calculation includes the contribution of the Hayward term, while that of \cite{Dong16} follows a method similar to \cite{MaldaAitor}, where one substitutes a neighborhood of $\Gamma$ by a thin tube around it, and then considers the variation (with respect to $n$) of \eqref{Ihat-saddle}. The Hayward term clearly plays no role in such a construction.

\section{The gravitational modular flow}

Let us show that the Hayward term may explain the presence of the area operator in the modular Hamiltonian $K$ of gravity and the JLMS proposal \cite{JLMS}. This is a relation holding beyond expectation values and at an operator level, see e.g. \cite{Dong18-2} for related discussion.  

The important object is the generator of the (gravitational) modular flow, namely $\rho^{is}$ where $s$ is a real parameter, but it can be hard to compute directly in gravity. A way to do this is take advantage of the replica calculus of \be\label{modularflow}\rho^n = U(i 2\pi n)\ee in the bulk studied in the previous sections, considering the analytical extension of the modular parameter: $is \to n$ (e.g. see \cite{CHM}).

The modular flow satisfies basic properties of symmetry and the Kubo-Martin-Schwinger (KMS) condition, and the modular Hamiltonian in general QFT is the \emph{generator} of the modular flow \eqref{modularflow} that can be computed by the formula
\be\label{modularKqft}
K = -\, \lim_{n\to 0}\, U(-i2\pi n)\frac{\partial }{\partial n} U(i2\pi n) \, =  -\, \lim_{n\to 0}\, \rho^{-n}\frac{\partial \rho^{n}}{\partial n} 
\ee
Having into account that the density matrix in the boundary field theory can be approximated by the most probable representation 
$\rho(A) \approx \rho(B, \Gamma_{min}) $ in gravity, we can use expression \eqref{rhoB-sumGamma} (or \eqref{rhoB}) for $\beta =2\pi n$, such that the bulk action is valued on a geometry $M_n$ 
\footnote{However the symmetry $Z_n$ is not exactly valid here because of the arbitrary conditions $h^\pm $ on the branches $B^\pm$, that deforms the bulk geometry, so \eqref{cociente-replica} is only approximation as this effect is negligible. This would be the case, for instance, if the $\Gamma$ is near the asymptotic boundary.}. Then if  $a(\Gamma_{min}, \gamma)$ in these expressions is interpreted as operator upon quantization (see Sec. 3)\cite{Jafferis},
and using  \eqref{modularKqft}, results the modular Hamiltonian
\be\label{KA}
K(A) = \frac{a(\Gamma_{min}, \gamma  )}{4G} + K_{bulk}(B)
\ee
where $K_{bulk}(B)$ is the modular Hamiltonian of the entanglement wedge $B$. 
In special cases with $U(1)$ symmetry, e.g. a black hole in the vacuum state ($\lambda \equiv 0$), $K_{bulk}$ coincides with the canonical Hamiltonian of the bulk theory. All these cases can be related (via the CHM map) to spherical entangling surfaces $\partial A$ on the boundary theory \cite{CHM}.
This result shows that the gravitational modular flow contains the area operator and reproduces the JLMS formula. 

\section{Conclusions}

In this work 
an area prescription for the holographic Rényi entropies in a purely gravitational formulation is presented. We have shown that the area term, which is usually computed through an auxiliary back-reacting codimension-2 brane, follows from including a \emph{necessary} boundary (Hayward \cite{Hayward93}) term in the bulk action for the 
geometries built from global pure states represented in Fig \ref{fig:H1}b.
In particular we established clearly the relation of the $n$-th Rényi entropy with the solution of $n$ consecutive copies of the boundary conditions on $M_n$,
recovering the holographic prescriptions for both $S_n$ and $\hat S_n$ \cite{Dong16,Dong18,Takayanagi19}, additionally shedding light on the origin of the area operator present in the modular Hamiltonian \cite{ensayoarea, Jafferis,JLMS}.
In fact our approach manifestly includes a term with matrix elements of the area operator, which would be difficult to explain from formulations without a Hayward term. 

Specifically, we considered a holographic CFT density matrix built via an Euclidean wave-function, possibly coupled to external sources such that it describes an excited state \cite{us1,Marolf,us5}. 
We project the state in a basis of two smooth regions corresponding to subregions $B$ and $\bar B$ glued together by a codim-2 surface $\Gamma$ that imposes a fixed $\pi-\beta/2$ deficit angle on the bulk $\Gamma$ splitting, where $\beta=2\pi n$ , see Fig. \ref{fig:H1}.  
The Hayward boundary term simplifies computations provide area laws for both $S_n$ and $\hat S_n$, albeit for different geometries\footnote{
A nice thermodynamical analogy of these two constructions can be made, where one defines a (micro)canonical description of the system either keeping fixed the (area element)deficit angle on $\Gamma$, corresponding to fixing the (energy)temperature in the CFT partition as thermodynamical variables. This complements recent discussion on the matter \cite{Dong18,Takayanagi19,Akers18}.}, described in Sec. 6. In the original holographic proposal \cite{Dong16}, an \emph{auxiliary} cosmic brane term is needed to produce the gravity solutions with conical singularity, such that the free energy is simply the (euclidean) gravitational on-shell action, but there is no contribution from the brane action itself. In contrast, the present formulation avoids this conceptual issue and capture both ingredients simultaneously: the total action is purely gravitational from the beginning, where the would-be brane action is nothing but the Hayward term, which \emph{is not} auxiliary in any sense but mandatory for a well defined variational problem; and moreover, it provides the main ($1/G$) contribution to the free energy and entropy.
Additionally, the analysis of Sec 2 about the Hilbert space decomposition of holographic bipartite systems, 
implies a sum over surfaces $\Gamma$ in the gravitational partition function that effectively turns it dynamical, and the Hayward term works as a Nambu-Goto action.
This provides an unified and systematic framework to describe holographic prescriptions on different measures of entanglement entropy, and modular hamiltonian 
with a term that can be interpreted as the \emph{area operator}. 
In the point of view adopted in this approach, the gravity edge modes associated to $\Gamma$ (and studied in Ref. \cite{Takayanagi19}) are a property of the basis where the state is projected, i.e. of the initial surface on which the set of field configurations describes a basis of the Hilbert space.

Although we have worked in the AdS/CFT framework for concreteness, this prescription allows to calculate the reduced density operator for any holographic field theory defined on the boundary  from the computation in the dual gravitational theory.
For instance, Einstein gravity without cosmological constant $\Lambda\equiv 0$ on spacetimes with an arbitrary (not necessarily asymptotic) boundary $\partial M$, such that the boundary condition is assumed to define some field theory on $\partial M$, and it is holographic in the sense that the gravitational theory can be interpreted as a suitable model to do approximations to the full calculations \cite{MaldaAitor}.

One could also envision other entanglement measures being holographically accounted for in terms of a Hayward term in a similar fashion as presented in this work. A covariant generalization of this construction, in the fashion of the HRT prescription \cite{HRT}, should also be possible. This would also involve an extremal surface ending on $\partial A$, but in the Lorentzian spacetime. One should thus extend this study in a complexified SvR-like \cite{SvRL} extension to the path integrals formulae. We leave this study for future research.

\subsection*{Acknowledgements }

The authors are specially indebted to Horacio Casini and Raul Arias for many fruitful discussions. Work supported by UNLP and CONICET grants X791, PIP 2017-1109 and PUE B\'usqueda de nueva F\'\i sica.  

\appendix

\section{Examples of conical geometries built with replicas}

For the sake of clarity, we present some simple examples of $3$d Euclidean spacetimes with deficit angle, proportional to the tension of the effective brane \cite{Vilenkin}, used in the main body of the text. These be can thought as higher dimensional metrics, with translations symmetry along the transverse codimension-2 surface $\Gamma$.

An example of manifold $M_n$ built in Sec. 6.1, with symmetry $U(1)$, is obtained through a $n$-times replicated BTZ solution with a consequent deficit angle of $2\pi(n-1)$:
\begin{equation}\label{metric-Mn}
ds^2 = r^2 d\tau^2 + \frac{dr^2}{r^2+1} + (r^2+1)dX^2\;~~~~~~;\qquad 0\leq \tau\leq 2n\pi 
\end{equation}
where $X$ denote the transverse coordinates. 
The minimal surface $\Gamma_{n, min}$ in this construction corresponds to $r=0$ has area 
\be a(M_n, \Gamma_{n, min})= a_1 \equiv \int dX \ee
which is independent on $n$ in this case. This is related to the fact that the quotient of this solution is $M_1$ which does not have conical singularity. This is essentially an example of the Fursaev's construction \cite{Fursaev}, and agrees with one of the results of  \cite{Dong18}.

On the other hand, a solution of  \eqref{I-hat}, $(\hat{M}_n ,\hat{\Gamma}_{n, min})$, is built such that it has a deficit angle $\hat \delta = 2\pi(n-1)/n$. The resulting manifold satisfies the original asymptotic boundary conditions but develops a conical singularity in the interior (see \cite{MaldaAitor})
\begin{equation}\label{metric-maldaaitor-smooth}
ds^2 = r^2 d\tau^2 + \frac{dr^2}{r^2+\frac{1}{n^2}} + \left( r^2+\frac{1}{n^2}\right) dX^2\;~~~~~~~~;\qquad  0\leq \tau\leq 2\pi \;, 
\end{equation}
and the corresponding minimal area is \be a(\hat{M}_n ,\hat{\Gamma}_{n, min})= \frac{a_1}{n}\;\;.\ee
Observe that this solution can be obtained by taking the $Z_n$ quotient of a spacetime \emph{without} conical singularity, consisting of the same metric \eqref{metric-maldaaitor-smooth} but such that the range of $\tau$ is $[0, 2\pi n]$.

\end{document}